\documentstyle[12pt]{article}
\newcommand{\be}{\begin{eqnarray}}
\newcommand{\en}{\end{eqnarray}}
\begin{document}
\begin{titlepage}
\setlength{\textwidth}{5.9in}
\begin{flushright}
EFI 98-27\\
MPI-Ph/98-52  \\
\end{flushright}
\begin{center}
\vskip 0.3truein
{\Large\bf {Reduction of Dual Theories }} \\
\vskip0.8truein
{Reinhard Oehme}
\footnote{E-mail: oehme@control.uchicago.edu}
\vskip0.5truein
{\it Enrico Fermi Institute and Department of Physics}\\
{\it University of Chicago} \\
{\it Chicago, Illinois, 60637, USA}
\footnote{Permanent Address}\\
{\it and}\\
{\it Max-Planck-Institut f\"{u}r Physik}\\
{\it - Werner-Heisenberg-Institut -}\\
{\it 80805 Munich, Germany}
\end{center}
\vskip0.5truein
\centerline{\bf Abstract}
\vskip0.3truein
In view of the presence of a superpotential, the dual
of a gauge theory like SQCD contains two coupling
parameters. The method of the Reduction of Couplings
is used in order to express the parameter $\lambda$
of the superpotential in terms of the dual gauge
coupling $g$. In the conformal window and above it, a unique,
isolated solution is obtained. It is given by $\lambda(g^2) = g^2 ~f $. 
Here $f$ is a function of the the number 
of colors and the number of flavors,
and it is known explicitly. This solution is valid to all orders in the 
asymptotic expansion, and it is the appropriate choice for the
dual theory. The same solution exists in the
free magnetic interval. A `general' solution with non-integer powers 
is discussed, as are some exceptional cases.
\end{titlepage}
\newpage
\baselineskip 18 pt
\pagestyle{plain}
\setlength{\textwidth}{5.9in}
\setcounter{equation}{0}
\vskip0.2truein
{\bf 1. Introduction}

In earlier publications \cite{DSP,NVS},
we have shown that certain results
about the phase structure
of SQCD and similar supersymmetric gauge theories,
which are obtained with the help
of duality and holomorphy \cite{WIS,SEI,SEN},
are in {\it quantitative}
agreement with the consequences
of our earlier work \cite{ROC,RLP,LOP}
involving analyticity and
superconvergence of the gauge field
propagator \cite{WZS,ROS,OWG}, \cite{NIC,NIP}.
This connection is of considerable interest,
because the superconvergence
arguments are also valid for non-supersymmetric theories.
A particular problem in this comparison,
and generally in the application of
duality, is the more detailed characterization
of the dual theory beyond the anomaly
matching conditions. In \cite{DSP}, we have already
given a brief sketch of the use,
for this purpose, of the method of reduction of couplings
\cite{OZS, PRO, WZH}. In dual SQCD, for
example, the reduction method makes it possible to
express the Yukawa
coupling of the superpotential as a function
of the coupling paramter associated with
the dual gauge group. The result is a
one-parameter magnetic theory, dual to
the one-parameter electric gauge theory.

A priori, the reduction equations relate the original
dual theory with two coupling parameters to a set of
solutions which also has two parameters. There are two
solutions with only the primary gauge coupling and
no other parameters. They
give rise to theories with remormalized power series
expansions of their Green's functions. Asymptotically
associated with one of these solutions is a 
`general' solution. It contains a free parameter besides
the gauge coupling, and generally leads to asymptotic
expansions with fractional powers of this coupling.
These expansions do not correspond to renormalized
perturbation theory in the usual sense.

Up to this point, only the renormalization group has
been used. However, one of the power series solutions
is excluded by the requirements of duality, like
preservation of global symmetries. The general solution
does not result in a conventional renormalizable theory.
Hence, there remains one single-coupling theory
with renormalized asymptotic power series expansion,
which is the appropriate dual of the original SQCD.

As we will see,
the reduction brings out essential features of the
dual theory which are not apparent in the two-coupling
formulation.

It is the purpose of this paper, to present the results 
of the reduction of couplings in the conformal window,
where the original and the dual theory are
asymptotically free at small distances. In addition, we
perform the reduction in the free magnetic region,
where the correlation functions at large
distances are those of the infrared-free
magnetic gauge theory.
We consider SQCD with the gauge group $G = SU(N_C)$
and N = 1 supersymmetry. There are $N_F$ superfields $Q_i$
and their antifields ${\tilde Q}^i ~,~
i = 1,2,...,N_F$ in the fundamental representation.
These fields are assumed to be massless. Otherwise, we suppose
that there is a mass-independent renormalization scheme
leading to renormalization group coefficients which
are independent of mass parameters.
The generalization to other
gauge groups is certainly possible \cite{ROP}\cite{MOT}. 
In the presence of matter
superfields in the adjoint representation,
the construction of dual theories requires
a superpotential also on the electric
side \cite{KUT}, and a corresponding
reduction would be indicated already there.
For a discussion of duality in general superconformal N = 1 models,
we refer to \cite{KLZ}.

\vskip0.2truein 

{\bf 2. Reduction}

In order to fix the notation, and for the
later discussion of fixed points
in the conformal window, we reproduce the one- and
two-loop $\beta$-function
coefficients for SQCD:
\begin{eqnarray}
\beta_e (g_e^2)~=~\beta_{e0}~g_e^4~+~\beta_{e1}~g_e^6~+~ \cdots ,
\label{1}
\end{eqnarray}
with
\begin{eqnarray}
\beta_{e0}
~&=&~(16\pi^2)^{-1} (-3N_C ~+~ N_F ) \cr
\beta_{e1}~&=&~(16\pi^2)^{-2} \left( 2N_C(-3N_C + N_F) ~+~
4N_F \frac{N_C^2 - 1}{2N_C} \right) ~.
\label{2}
\end{eqnarray}
The label $e$ indicates that these $\beta$-functions refer to the `electric'
theory. Later, we will omit a corresponding label for the functions of the
`magnetic' theory.
The theory dual to SQCD involves the gauge group
${G^d} = SU({N^d_C})$, with ${N^d_C} = N_F - N_C $. There
are $N_F$ quark superfields ${q_i}, ~{\tilde q}^i,  ~i =
1,2,...,N_F$
in the fundamental representation of ${G^d}$, as well as $N^2_F$ gauge
singlet superfields $M^i_j$ . The superfields $M$ are independent,
and cannot be constructed from $q$ and $\tilde q$.
The number of flavors $N_F$ is the same for
SQCD and for dual SQCD, because both theories must have equal global
symmetries.
The construction of the dual theory is done essentially on the basis of the
anomaly matching conditions \cite{SEN,IOS},
which require the colorless fields  $M^i_j$
and their coupling via the superpotential
\begin{eqnarray}
W ~= ~ \sqrt{\lambda} M^i_j q_i {\tilde q}^j.
\label{3}
\end{eqnarray}
A priori, the dual theory has two  parameters, the
gauge coupling $g^2$ and the Yukawa coupling
$\lambda$. The presence of the superpotential is of importance, 
not only for the coupling of the $M$-superfield, but also in order
to remove a global U(1) symmetry acting on this field, which would
otherwise be present. This symmetry has no counterpart in the
electric theory. It would destroy the match between the physical
symmetries of both theories, which is essential for duality.
Consequently, we will accept only reductions which do not switch-off
the superpotential. 

We write the $\beta$-functions \cite{AND} of the magnetic theory
in the form:
\begin{eqnarray}
\beta (g^2, \lambda)~&=&~\beta_0~g^4~+~(\beta_1~g^6~+~
\beta_{1,\lambda}~g^4\lambda) ~+~\cdots  \cr
\beta_\lambda (g^2, \lambda)~&=&~c_\lambda g^2 \lambda ~+~
c_{\lambda\lambda} \lambda^2 ~+~\cdots ~.
\label{4}
\end{eqnarray}
The coefficients are given by
\begin{eqnarray}
\beta_0~&=&~(16\pi^2)^{-1} (3N_C - 2N_F ) \cr
\beta_1~&=&~(16\pi^2)^{-2} \left( 2(N_F - N_C)(3N_C - 2N_F) ~+~
4N_F \frac{(N_F- N_C)^2 - 1}{2(N_F - N_C)} \right) \cr
\beta_{1,\lambda}~&=&~(16\pi^2)^{-2} \left( -2N^2_F \right)  \cr
c_\lambda~&=&~(16\pi^2)^{-1} \left( -4 \frac{(N_F - N_C)^2 - 1}{2(N_F -
N_C)} \right) \cr
c_{\lambda\lambda}~&=&~(16\pi^2)^{-1} \left(3N_F - N_C) \right) ~.
\label{5}
\end{eqnarray}

We now want to express the Yukawa coupling $\lambda$ as a function of the
gauge coupling $g^2$, which we choose as the primary coupling parameter:
$\lambda = \lambda(g^2)$ . The method of reduction
is based upon the requirement that the Green's functions of the reduced
one-parameter theory satisfy the appropriate renormalization group equations
involving the single coupling parameter $g^2$. The corresponding
$\beta$-function
is then given by
\begin{eqnarray}
\beta (g^2) ~=~ \beta (g^2, \lambda(g^2)) .
\label{6}
\end{eqnarray}
Comparing the renormalization group relations for the one-paramter and the
two-parameter theories, we obtain the {\it reduction equations}
\footnote{For recent surveys of the reduction method, see
\cite{PRO,RIN}. The case of two couplings has been discussed in
detail in \cite{OZC}.}
\begin{eqnarray}
\beta (g^2) \frac{d \lambda(g^2)}{d g^2} ~=~
\beta_\lambda (g^2)~,
\label{7}
\end{eqnarray}
where $\beta (g^2)$ has been defined in Eq.(6) above, and
\begin{eqnarray}
\beta_\lambda (g^2) ~=~ \beta_\lambda (g^2, \lambda(g^2))~.
\label{8}
\end{eqnarray}
The reduction equation (7) is necessary and sufficient for the validity of
the
renormalization group equations for the Green's functions
of the reduced one-parameter theory. It
can also be obtained from the renormalization group equations for the
effective couplings $ \bar{g}^2(u)$ and $\bar{\lambda}(u)$ by elimination
of the scaling parameter $u$ in favor of the the function $\lambda(u)$,
using proper caution.
With the asymptotic expansions of the $\beta$-functions as given in Eq.(4),
the reduction equation is singular at $g^2 = 0$. Uniformisation
transformations
can remove the singularity and show that all solution have asymptotic series
expansions at the origin \cite{OZS,WZG}. These expansions may contain
non-integer powers.

After these preliminaries, we return to the reduction of dual SQCD. At first
we
consider solutions \cite{OZC,WZZ,ONA}  which have asymptotic power
series expansions
for $g^2 \rightarrow 0$.  We can restrict ourselves to
solutions where the ratio $\lambda(g^2)/g^2$ is bounded at $g^2=0$. 
As seen from the reduction equation (7), an Ansatz with $\lambda(0) \neq 0$
would require $\beta_{\lambda}(0, \lambda(0)) = 0$ since we have
$\beta(0, \lambda(0)) = 0$. This constraint
is generally not fullfilled.

We write
\be
\lambda(g^2) = g^2 f(g^2)~, ~\mbox{with}~~ f(g^2) = f_0 +
\sum_{m=1}^{\infty}
\chi^{(m)} g^{2m} ~.
\label{9}
\en
Substitution into the reduction equation (7) yields the fundamental one-loop
relation
\be
\beta_0 f^0 ~=~ \left( c_{\lambda\lambda} f^0 ~+~
c_\lambda \right) f^0 ~.
\label{10}
\en
There are two solutions:
\be
f^0 = f_{00} = 0~~ \mbox{and}  ~~f^0 = f_{01} =
\frac{\beta_0 - c_\lambda}{c_{\lambda\lambda}} ~,
\label{11}
\en
where $f_{01}$ is a function of $N_C$ and $N_F$, and is given by
\be
f_{01}(N_C, N_F)~=~\frac{N_C\left( N_F - N_C - {2}/{N_C} \right)}
{(N_F - N_C)(3N_F - N_C)} ~.
\label{12}
\en
Here and in the following, we do not consider possible additional
terms which vanish exponentially or faster \cite{ONA}.

The one-loop equation (10) is the fundamental relation for
reductions.  One loop criteria also decide whether the higher order
coefficients are determined by the reduction equation.  Up to
$m+1$ loops, we have the relations:
\be
\left(M (f^0) - m \beta_{0} \right)
\chi^{(m)} = \left(\beta_m (f^0) f^0 -
\beta^{(m)} (f^0)\right) + X^{(m)} ,
\label{13}
\en
where $m=1,2,\ldots ~$.
Here we have written the expansions of $\beta(g^2)$ and
$\beta_{\lambda}(g^2)$
in a form, which will also turn out to be very useful later:
\be
\beta (g^2) = \beta (g^2, g^2 f (g^2)) = \sum^\infty_{n=0}
\beta_{n} (f) (g^2)^{n+2} ~ \cr
\beta_{\lambda} (g^2) = \beta_{\lambda} (g^2, g^2 f (g^2)) =
\sum^\infty_{n=0}
\beta^{(n)}_{\lambda} (f) (g^2)^{n+2} ,
\label{14}
\en
where
\be
\beta_0 (f) = \beta_0 ,~~ \beta_1(f) = \beta_1 + \beta_{1\lambda} f ,~~
\beta_{\lambda}^{(0)} = c_{\lambda} f  +  c_{\lambda\lambda}  f^2 ,  ~~etc.
~.
\label{15}
\en
The coefficient $M(f^0)$ in Eq.(13) given by
\be
M (f^0) = c_{\lambda} + 2 c_{\lambda\lambda}
f^0 -
\beta_{0}~.
\label{16}
\en
The rest term $X^{(m)}$ depends only upon the coefficients
$\chi^{(1)},\ldots,\chi^{(m-1)}$, and upon the $\beta$--function
coefficients in Eqs.(14) of the order $m-1$ and lower,
evaluated at $f = f^0$. It vanishes for
$\chi^{(1)} = \ldots = \chi^{(m-1)} = 0$.
We see that the {\em one--loop} criteria
\be
\left( M(f^0)- m \beta_{0}\right)
 \not = 0
   ~~ for ~~m=1,2,\ldots
\label{17}
\en
are sufficient to insure that all coefficients $\chi^{(m)}$ in the
expansion (9) are determined.  Then the reduced theory has a
renormalized power series expansion in $g^2$.  All possible solutions
of this kind are fixed by the one--loop equation (10) for
$f^0$.
In order to discuss the solutions for dual SQCD, it is convenient to
consider characteristic intervals in $N_F$ separately.

\vskip0.2truein

{\bf 3. Conformal Window}
 
We consider first the {\it conformal window} where
{\it $\frac{3}{2} N_C ~<~ N_F ~<~3N_C $}.
\footnote{See page 450 of \cite{LOP}, where the existence of a phase
transition at $N_F = \frac{3}{2}N_C$ has already been derived
on the basis of superconvergence relations, and
\cite{ROC} for the corresponding non-SUSY result.}
Here both SQCD and dual SQCD are
asymptotically free at small distances, as seen from Eqs.(2) and (5).
We consider first the solution with $f^0 = f_{01}$ as given by Eq.(12).
In the widow we have $f_{01} > 0$, as required by the superpotential.
The factor in front of the coefficient $\chi^{(m)}$ in Eq.(13) is of
the form
\be
\left( M(f_{01}) - m\beta_0 \right) ~=~ -\beta_0 (\xi + m) ~,
\label{18}
\en
where
\be
\xi(N_C, N_F)~=~\frac{N_C\left( N_F - N_C - {2}/{N_C} \right)}
{(N_F - N_C)(2N_F - 3N_C)} ~.
\label{19}
\en
In the widow, we have $\xi > 0$, and hence the coefficients $\chi^{(m)}$
in the expansion (9) are all determined.

To go further, it is useful to consider
{\it regular reparametrisations} of the theory \cite{OZC}.
These transformations leave the physics unchanged, but the 
$\beta$-functions of the original theory are 
generally not invariant. 
An important exception are the lowest order terms. We will use
reparametrization
in order to transform the fixed coefficients $\chi^{(m)}$ to zero. The
transformations in question are of the form
\be
g'^2 ~=~ g^2 + a^{(20)} g^4 + a^{(11)} g^2 \lambda +\cdots , \cr
{\lambda}' ~=~ \lambda + b^{(20)} {\lambda}^2 + b^{(11)} \lambda g^2 +
\cdots  ~,
\label{20}
\en
and they leave the one-loop quantities $\beta_0$ ,~
$\beta_{\lambda}^{(0)}(f^0)$,~
$f^0$ ~ and $M(f^0)$  invariant. On the other hand, there is a
sufficient
number of free parameters in the transformations (20), so that we can arrange
for all coefficients ${\chi}^{(m)}$ in the expansion (9) to be transformed
to zero.
Hence the power series solution (9) reduces to the simple form
\be
\lambda (g^2) ~=~ g^2 ~f_{01}(N_C, N_F)~,
\label{21}
\en
with $f_{01}$ given by Eq.(12). The $\beta$-functions of the reduced theory,
as defined by the solution (21), are now simply given by Eqs.(14) with the
argument $f$ of the coefficient functions  replaced by $f_{01}(N_C N_F)$ ,
so
that
they are constants:
\be
\beta (g^2) = \beta (g^2, g^2 f_{01}) = \sum^\infty_{n=0}
\beta_{n} (f_{01}) (g^2)^{n+2}  ~, ~~
\beta_{\lambda} (g^2) = f_{01} \beta (g^2) ~.
\label{22}
\en
The second relation follows from the reduction equation (7) with Eq.(21).
The coefficient $\beta_0$ is as given in Eq.(5), and for $\beta_1 (f_{01})$
we obtain explicitly

\newpage
\be
(16\pi^2)^2 \beta_1(f_{01}) ~=~   2(N_F - N_C)(3N_C-2N_F) ~+~
4N_F \frac{(N_F-N_C)^2-1}{2(N_F-N_C)}  \cr
- 4N_F^2 ~\frac{N_C (N_F-N_C-2/N_C)}
{2(N_F-N_C)(3N_F-N_C)} ~.
\label{23}
\en
These relations are used later in connection with the infrared fixed point
of dual SQCD in the conformal window near $N_F = \frac{3}{2}N_C$.
We must note here, that for the expansion (22), in addition to $\beta_0$,
the two-loop coefficient $\beta_1(f_{01})$ is {\it reparametrization invariant}.
This result follows because $f_{01}$ satisfies the reduction equation
(10) \cite{ONA}.

We now turn to the second solution of the reduction equation
(7), which is associated with the one-loop result $f_0 = f_{00} = 0$
in Eq.(11). Here the expansion coefficients $\chi^{(m)}$ in the series
(9) have the factor
\be
(M(0) -m\beta_0) ~=~ +\beta_0 (\xi - m)  ~,
\label{24}
\en
with $\xi (N_C, N_F) $  given in Eq.(19). In all cases where $\xi$, which is
positive in the conformal window, is not an integer, the coefficients
$\chi^{(m)}$ are again determined. They all vanish, as seen from Eq.(13),
and
hence the corresponding solution is
\be
\lambda(g^2) ~\equiv~ 0 ~.
\label{25}
\en
This solution represents a well defined renormalized theory with an
asymptotic power series expansion in the gauge coupling $g^2$ .
As we have discussed before, it is not acceptable as a dual theory
of SQCD because of the vanishing superpotential.

The situation described above  prevails for most values 
of $N_C$ and $N_F$ in the window.
An exception is the case $N_C = 3$, $N_F = 5$, where we have $\xi (3, 5) =
+2$.
Then the coefficient of the power $g^4$ in the expansion (9) vanishes, and
after reparametrization, we have the solution
\be
\lambda(g^2) ~=~ A g^6 + {\chi}^{(3)} g^8 + \cdots .
\label{26}
\en
Here the coefficient $A$ is undetermined. 
Once $A$ is given, the higher order
coefficients are fixed. For $A = 0$, they all vanish and we
have again Eq.(25). For the exceptional situation mentioned here, 
and two similar cases encountered later in the free magnetic
interval, the dual gauge group is ${G^d} = SU(2)$, which has 
some special features not present for lager values of $N^d_C$.
Here we
consider the solution (26) as a special case of the `general' 
solution discussed below.

It remains to discuss the `general' \cite{OZC,ONA}
solutions of the reduction equation in
the conformal window. They generally involve
non-integer powers of $g^2$. Trying an
Ansatz of this kind, we see that there are no such solutions associated with
$f^0 = f_{01}$ in Eq.(11). On the other hand, for $f^0 = f_{00} = 0$, 
we have a solution of the form
\be
\lambda (g^2) ~=~ B g^{2 + 2\xi} + \cdots ,
\label{27}
\en
where $\xi$ is again given by Eq.(19). It is non-integer in the 
window, at least for $N_C < 16$. An exeption is the
case $N_C = 3$, $N_F = 5$ discussed above, where the dual gauge 
group is $SU(2)$. Higher order terms are of
the form $(g^2)^{n \xi + m}$, with appropriate distinct powers. 
The coefficient $B$ is undetermined. If $B$ is
fixed, the coefficients of the higher terms are also determined, and
they vanish if $B = 0$. 

It is important to realize, that the two coupling parameters of
the original version of the dual theory are reflected, in the set
of solutions for the reduction equations, in the free coefficient
$B$ of the `general' solution (27) and the reaining primary
coupling $g$. Within the set of solutions, the power series 
$\lambda(g^2) = g^2 f_{01} $ stand out as being appropriate 
for duality. The other power series solution $\lambda(g^2) \equiv 0$
is excluded. The `general' solution is associated with this 
forbidden solution, since both approach each other asymptotically.
For these reasons, and because of the lack of standard asymptotic
power series expansions for the Green's functions, 
we do not consider theories involving
the `general' solution as appropriate duals to SQCD.
\footnote {Although they would not be appropriate for dual 
theories with renormalized asymptotic expansions,
one may ask about the possibility of other `general' solutions
of the reduction equation, which do not approach the power
series (21) or (25). If such solutions should
exist, they would require more explicite, non-asymptotic 
knowledge of the $\beta$-functions. However, one can use the 
theorems of Ljapunov \cite{LJA}, as generalized by Malkin 
\cite{MAL}, in order to obtain information 
about the possible existence of such solutions
on the basis of the linear part of the differential equation
\cite{OSZ,ONA}. In a finite neighborhood of the origin, 
no solutions of this kind are expected for the theories
considered here. Stability or instability of the solutions
(21) and (25), with $f(g^2)$ approaching $f^0$, is determined
by the positive or negative sign of $\beta^{-1}_0 M(f^0)$
respectively, as has been discussed above. } 

Since $\xi > 0$ in the window, the Yukawa coupling in (27)
vanishes faster than $g^2$, paticularly near the lower end 
of the widow considered here, where $\xi$ becomes very large.  
This behavior would have implications for the infrared fixed-point, as will
be discussed below.

We conclude that, in the conformal window, the solution
$\lambda (g^2) = g^2 f_{01} $ represents the unique and isolated
one-coupling theory which is dual to SQCD. It is isolated or `unstable' \cite{ONA},
because there are no other solutions approaching it for $g^2 \rightarrow 0$.
The theory has an asymptotic power series expansion, and the one-loop
character of the anomalies is preserved.

\vskip0.1truein
{\it Infrared Fixed-Points}\\
Since a very long time, it is well known that QCD and SQCD appear to have
non-trivial infrared fixed points for values of $N_F$ near the point
where asymptotic freedom is lost \cite{BZK}. We are interested in possible fixed
points in the window
$\frac{3}{2} N_C ~<~ N_F ~<~3N_C $ for the two theories considered
here, which are dual to each other. Very near the point $N_F = 3N_C$,
approaching from below, the electric theory is weakly coupled at large
distances, and we may
use the one- and two-loop $\beta$-function coefficients in Eq.(2) to find
\be
\beta_e (g_e^{*2}) ~=~  0 ~~~~~\mbox{for}~~~~~ \frac {g_e^{*2}}{16\pi^2} 
~\approx~ \frac{3N_C-N_F}
{6(N_C^2-1)} ~ .
\label{28}
\en
As mentioned, we have assumed that $(3N_C - N_F)$ is positive and sufficiently
small compared to $N_C$ in order to get a small value of 
$ {g_e^{*2}}/{16\pi^2}$. We have neglected here all higher order terms
in $g_e^{*2}$, and evaluated the coefficient of  $(3N_C - N_F)$ at
$N_F = 3N_C$.

At the other end of the window, very near the point $N_F = \frac{3}{2} N_C$,
the magnetic theory is weakly coupled in the infrared. We can use the
unique
reduced
theory, as defined by Eq.(21), in order to obtain the fixed point there.
With
the
$\beta$-function coefficients in Eqs.(5) and (23), we find
\be
\beta (g^{*2})  ~=~ 0  ~~~~~\mbox{for}~~~~~ \frac {g^{*2}}{16\pi^2} 
~\approx~ \frac{7}{3} ~\frac{N_F-\frac{3}{2} N_C}
{\frac{N_C^2}{4}-1} ~,
\label{29}
\en
with assumptions analogous to those described above, but now referring 
to $(N_F-\frac{3}{2} N_C)$. Certainly, larger values of $N_C$ are needed
for this approximation to be useful.

It is relevant here, and has been pointed out before, that 
the two-loop coefficient $\beta_1(f_{01})$ is {\it reparametrization invariant}
\cite{ONA}, as is $\beta_0$. 

The important proposal by Seiberg \cite{SEI,SEN} is, that for given
values of $N_C$ and $N_F$ in the conformal window, 
the electric and the magnetic
theories flow to the {\it same} fixed point. For example, near 
the lower end of the window, with $N_F$ near $\frac{3}{2}N_C$, 
the magnetic theory is weakly coupled in the
infrared, and we find that a fixed point is present. In contrast,
the electric theory is strongly coupled there. Since both theories
should be the same at the fixed point, we can obtain information about
the electric theory by using the weakly coupled magnetic dual.

Although we did not consider the `general' solution (27) to be appropriate 
for the dual theory, it
may be of interest to note, that with this solution, as well as
with the exceptional solution (26), the corresponding reduced theories
flow to fixed points which are different
from those of the theory with $\lambda(g^2) = g^2 f_{01}$. 
For $({N_F}/{N_C} - {3}/{2})$ sufficiently
small, where we can obtain a crude estimate, these fixed points are 
more near to the
the Coulomb phase fixed point of the theory with $\lambda(g^2) \equiv 0$.  

\vskip0.1truein
{\it $ N_F > 3N_C$} \\
We add here some remarks about the region above the window.
For these values of $N_F$, the magnetic theory is asymtotically free at small
distances. We again have $f_{01} > 0$ and $\xi > 0$ , as seen from Eqs.(12)
and (19). The results of the reduction method are the same as in the
conformal window discussed above, with the reparametrized power
series solution given in Eq.(21) representing the dual one-parameter
theory. In contrast to the situation at the lower end of the conformal
window, for the larger values of $N_F$ considered here, the magnetic theory
is strongly coupled, and the spectrum of the theory is that of the electric
Lagrangian. It is the electric theory which is infrared free in the
region of $N_F$ considered here.

\vskip0.2truein

{\bf 4. Free Magnetic Phase}

The {\it free magnetic phase} is the interval
{\it $N_C + 2 ~\leq~ N_F ~<~ \frac{3}{2}N_C $}.
It is non-empty
for $N_C > 4$, which we assume in the following. For convenience, we
consider first $N_F > N_C + 2$, leaving the boundary case for later.
The electric theory is asymptotically free at small distances, but
strongly coupled otherwise, while the magnetic theory is infrared
free. Hence, at low energies, the spectrum of the theory is that of the
magnetic Lagrangian. Although, in view of the
lack of asymptotic freedom at small distances,
the theory may not exist as a strictly local field theory, it can be
considered as a large distance limit of an appropriate
brane construction in superstring theory,
which can also reaffirm duality \cite{KUS}. 
The same remarks apply to the electric theory
discussed above in the region  $N_F > 3N_C$.

For the reduction of the free magnetic theory, we consider again the
asymptotic expansion for $g^2 \rightarrow 0$. Now however, this
limit corresponds to an approach to the {\it trivial infrared fixed
point}. The solutions of the reduction equation are similar to those
we have discussed above in the conformal window.
The important change is, that the function $\xi(N_C, N_F)$ defined in
Eq.(19) is now negative,
while the coefficient $f_{01}(N_C, N_F)$  in Eq.(12) remains positive,
as required by the reality condition for the superpotential (3).
Consider first the power series solution in Eq.(9) associated with the
one-loop coefficient $f_{01}$ from Eq.(11). In calculating the expansion
coefficients $\chi^{(m)}$ using Eq.(13), we now have the factor
\be
\left( M(f_{01}) - m\beta_0 \right) ~=~ -\beta_0 (\xi + m) ~,
\label{30}
\en
which could possibly vanish. But inspection of $\xi(N_C, N_F)$ given in
Eq.(19) shows, that it is not a negative integer in the free magnetic
region considered here, at least for $N_C < 16$. 
Hence we have again a well determined power
series solution. Using the regular reparametrizations (20), it can be
transformed into the simple form given
in Eq.(21):  $\lambda (g^2) = g^2 ~f_{01}$.

With the other solution of the one-loop euqation, $f^0 = f_{00} = 0$, 
we find for the coefficient of $\chi^{(m)}$ in Eq.(13)
\be
\left( M(0) - m\beta_0 \right) ~=~ -\beta_0 (-\xi + m) ~,
\label{31}
\en
which cannot vanish since $\xi < 0$ in the free magnetic region.
Hence all the coefficients $\chi^{(m)}$ are determined,
and it follows from Eq. (13) that they all vanish, so that
the solution $\lambda (g^2) \equiv 0$
is obtained. As we have pointed out before, because the presence 
of the superpotential (3) is required, 
this solution cannot be used
as the magnetic theory.

We finally consider the `general' solution of the reduction equation
for the free magnetic region. It is easily seen to be of the
form
\be
\lambda (g^2) ~=~ g^2 f_{01} ~+~ C ~g^{2 + 2|\xi|} ~+~ \cdots ~,
\label{32}
\en
with an undetermined coefficient $C$
and $|\xi| > 1$. In Eq.(32), we have implied
a reparametrization transformation in order remove possible integer
powers below the  $C$-term. Higher order terms involve distinct
powers of the form $(g^2)^{n|\xi| + m}$ with appropriate integers
n and m. Their coefficients are determined once $C$ is fixed,
and the choice $C = 0$ leads back to the solution (21).
In contrast to the situation in the conformal window, we see that here
the power series solution (21) is stable in the sense that
the general solution (32) approaches it near $g^2 = 0$ . On the other hand, the
null solution (25) is isolated or unstable.

As we have discussed in the case of the conformal window, the 
`general' solution (32) would not give rise to a dual theory with a
renormalized asymptotic power series expansion. Cosequently.
there remains the special power series solution (21)
which should be used as the magnetic
theory in the free magnetic region considered. The one-loop character
of the matching conditions is certainly valid for this solution.
In addition, it connects with the solution appropriate for the
conformal window.


It remains to mention the boundary case
{\it $N_F = N_C + 2$}. It still
belongs to the free magnetic phase provided $N_C > 4$, so that
$N_C + 2 < \frac{3}{2} N_C$. The magnetic gauge group $G^d$
is $SU(2)$, and there is some extra flavor symmetry.

For $N_F = N_C + 2$ , the results of the reduction
method are essentially the same
as described above for the free magnetic region 
with $N_F > N + 2$. There are two exceptions for
the power series solution (9) with $f^0 = f_{01}$, namely
$N_C = 5$ and $N_C = 7$.  For these cases the corresponding
coefficients of $\chi^{(m)}$ in Eq.(13) vanish, because
\be
\xi (N_C, N_F = N_C + 2) ~=~ - 1 - \frac{3}{N_C - 4}
\label{33}
\en
implies that $\xi (5, 7) = - 4$ as well as $\xi (7, 9) = - 2$ . For these
values of $N_C$, we do not have the completely determined special solution
(21) because of the undefined coefficient in the expansion (9). Rather
there are the expressions
\be
\lambda (g^2) ~=~ g^2 f_{01} ~+~ D ~g^{2 + 2|\xi|} ~+~ \cdots ~,
\label{34}
\en
with $|\xi| = 4$  and $|\xi| = 2$ respectively. As in general solutions
considered earlier, the coefficient $D$ is undetermined, and all later
terms are fixed once $D$ is given, or vanish if we choose $D = 0$.
The three exceptional cases we have encountered, all have $G^d = SU(2)$
as the dual gauge group, and there are special features we plan to
discuss elsewhere.

For smaller values of $N_F$, like $N_F = N_C + 1 , ~N_C$ , there is no dual
magnetic gauge theory. The spectrum is described by gauge invariant fields,
massles baryons and mesons, which are then the elementary magnetic quanta.

\vskip0.1truein
{\it $N_F ~=~ \frac{3}{2}N_C $}\\
At this important transition point of the magnetic theory we
have $\beta_0 = 0$, so that the function $\xi(N_C, N_F)$, given
in Eq.(19), is not defined. There are no general solutions, and
we have only the two power series solutions, which can be brought into
the forms (21) and (25) respectively. Again then the solution (21):
$\lambda (g^2) = g^2 f_{01}$ is the only choice for the magnetic
theory. Here the coefficient $f_{01}$, as defined in Eq.(12), is
given by
\be
f_{01} (N_C, N_F = \frac{3}{2} N_C) ~=~ \frac{2}{7} (1 - \frac{4}{N_C^2} ) .
\label{35}
\en
We have assumed $N_C > 2$ throughout.

\vskip0.2truein

{\bf 5. Conclusions }

We summerize the results of the reduction method:

In the {\it conformal window} $~~\frac{3}{2} N_C ~<~ N_F ~<~3N_C $ , at the
transition point  $N_F ~=~ \frac{3}{2} N_C $, as well as for
$N_F > 3N_C$, we have
the unique and isolated power series solution (21): ~
$\lambda (g^2) ~=~ g^2 f_{01}(N_C, N_F)$ , ~
which is valid to all orders in the asymptotic expansion in $g^2$.
Only possible contributions which vanish exponentially or
faster are not written.
It is the appropriate solution
for the dual magnetic theory associated with SQCD .

The same solution exists for the {\it free magnetic region} 
$~~N_C + 2 ~\leq~ N_F ~<~ \frac{3}{2}N_C $. There 
this solution is accompanied
by a `general' solution, which approaches it for $g^2 \rightarrow 0 $.
Except for the two special cases mentioned, 
the general solution has non-integer powers and a free parameter.
Since we want the magnetic theory to have asymptotic power
series expansions for the Green's functions,
we select the solution (21), which then
defines this theory for all $N_F > N_C + 2$. With the exception of
two values of $N_C$, we can also include the boundary case  $N_F = N_C + 2$.
In the free magnetic region, the asymptotic expansion is a
long distance expansion, because the magnetic theory is infrared-free,
while in the conformal window and above, we have asymptotic freedom
at small distances.

\vskip2.0truein
\centerline{ACKNOWLEDGMENTS}
I would like to thank Jisuke Kubo,
Klaus Sibold and  Wolfhart Zimmermann for very useful discussions.
It is a pleasure to thank Wolfhart Zimmermann, and the
theory group of the Max Planck Institut f\"ur Physik, - Werner Heisenberg
Institut -, for their kind hospitality in M\"unchen.
This work has been supported in part by the National Science Foundation,
 grant PHY 9600697.
\newpage


\begin{thebibliography}{99}
\bibitem{DSP}
{R. Oehme,  Phys. Lett. {\bf B399}, 67 (1997).}
\bibitem{NVS}
{R. Oehme,  `Supercovergence, Confinement and Duality', \\
{\it (Talk presented at the International Workshop on High Energy Physics,
Novy Svit, Crimea, September 1995)},
Proceedings, edited by G.V. Bugrij and L. Jenkovsky
(Bogoliubov Institute, Kiev, 1995) pp. 107-116;
hep-th/9511014.}
\bibitem{WIS}
{N. Seiberg and E. Witten, Nucl. Phys. {\bf B426}, 19 (1994) ; \\
ibid. {\bf B431}, 484 (1994). }
\bibitem{SEI}
{N. Seiberg, Phys. Rev. {\bf D49}, 6857 (1994). }
\bibitem{SEN}
{N. Seiberg, Nucl. Phys. {\bf B435}, 129 (1995). }
\bibitem{ROC}
{R. Oehme, Phys. Rev. {\bf D42}, 4209 (1990); ~
Phys. Lett. {\bf B155}, 60 (1987). }
\bibitem{RLP}
{R. Oehme, Phys. Lett. {\bf B232}, 489 (1989). }
\bibitem{LOP}
{R. Oehme, `Superconvergence, Supersymmetry and Conformal Invariance', \\
in {\it Leite Lopes Festschrift}, edited by
N. Fleury, S. Joffily, J. A. M. Sim\~{o}es, and A. Troper
(World Scientific, Singapore, 1988)  pp. 443 - 457; \\
University of Tokyo Report UT 527 (1988). }
\bibitem{WZS}
{R. Oehme and W. Zimmermann, Phys. Rev. {\bf D21}, 474, 1661 (1980). }
\bibitem{ROS}
{R. Oehme, Phys. Lett. {\bf B252}, 641 (1990).}
\bibitem{OWG}
{R. Oehme and W. Xu, Phys. Lett. {\bf B333}, 172 (1994); \\
ibid. {\bf B384}, 269 (1996). }
\bibitem{NIC}
{K. Nishijima, Prog. Theor. Phys. {\bf 75}, 22 (1986),
Nucl. Phys. {\bf B238}, 601 (1984); \\
K. Nishijima in {\it Symmetry in Nature}, Festschrift
for Luigi A. Radicati di Brozolo (Scuola Normale Superiore,
Pisa, 1989) pp. 627-655. }
\bibitem{NIP}
{K. Nishijima, Prog. Theor. Phys. {\bf 77}, 1053 (1987). }
\bibitem{OZS}
{R. Oehme and W. Zimmermann, Max Planck Institut
Report MPT-PAE/Pth 60/82 (1982), Commun. Math.
Phys. {\bf 97}, 569 (1985); R. Oehme, K. Sibold
and W. Zimmermann, Phys. Lett. {\bf B147}, 115 (1984);
W. Zimmermann,
Commun. Math. Phys. {\bf 97}, 211 (1985);
R. Oehme, K. Sibold and W. Zimmermann, Phys. Lett.
{\bf B153}, 142 (1985);
R. Oehme, Prog. Theor. Phys. Suppl. {\bf 86}, 
215 (1986).}
\bibitem{PRO}
{R. Oehme, `Reduction of Coupling Parameters', \\
{\it (Plenary talk at the XVIIIth International
Workshop on High Energy Physics
and Field Theory, June 1995, Moscow-Protvino, Russia)},
Proceedings, edited by V.A. Petrov, A.P. Samokhin and R.N. Rogalyov, 
pp. 251-270;
hep-th/9511006.}
\bibitem{WZH}
{W. Zimmermann, `Reduction in the Number of Coupling Parameters', \\
{\it (Talk given at the Ninth John Hopkins Workshop on Current Problems
in High Energy Particle Theory, June 1985)}, MPI-PAE/P-Th 44/85 .}
\bibitem{ROP}
{R. Oehme, (in preparation).}  
\bibitem{MOT}
{M. Tachibana, Phys. Rev. {\bf D58},
045015 (1998). }
\bibitem{KUT}
{D. Kutasov, Phys. Lett. {\bf B351}, 230 (1995);\\
D. Kutasov and A Schwimmer, Phys. Lett. {\bf B354}, 315 (1995);\\
K. Intriligator, R. Leigh and M. Strassler, Nucl. Phys.
{\bf B456}, 567 (1995). }
\bibitem{KLZ}
{A. Karsch, D. L\"ust and G. Zoupanos, 
Phys. Lett. {\bf B430}, 254 (1998); \\
A. Karsch, D. L\"ust and G. Zoupanos, hep-th/9711157; \\
A. Karsch, T. Kobayashi, J. Kubo and G. Zoupanos, hep-th/9808178.  }
\bibitem{IOS}
{K. Intriligator and N. Seiberg, Nucl. Phys. Proc. Suppl.
{\bf BC45}, 1 (1996),~
Nucl. Phys. {\bf B444}, 125 (1995). }
\bibitem{AND}
{ D.R.T. Jones, Nucl. Phys. {\bf B87}, 127 (1975);
R. Barbieri et al., Phys. Lett. {\bf B115}, 212 (1982);
A.J. Parkes and P.C. West, Nucl. Phys. {\bf B256}, 340 (1985);
P. West, Phys. Lett. {\bf B137}, 371 (1984);
D.R.T. Jones and L. Mezincescu, Phys. Lett. {\bf B136}, 293 (1984);
I.I. Kogan, M. Shifman, and A. Vainstein, Phys. Rev. {\bf D53},
4526 (1996). }
\bibitem{RIN}
{R. Oehme, `The Method of the Reduction of Couplings' , \\
{(\it Talk presented at the Ringberg Symposium on Quantum
Field Theory, Ringberg Castle, Tegernsee, June 1998)},
Proceedings, edited by P. Breitenlohner and D. Maison; (in preparation).}
\bibitem{OZC}
{R. Oehme and W. Zimmermann, Commun. Math.
Phys. {\bf 97}, 596 (1985). }
\bibitem{WZG}
{ W. Zimmermann, in the Vladimir Glaser Memorial Volume (1985), \\
MPI-PAE-PTH 85-17 (June 1985) .}
\bibitem{WZZ}
{W. Zimmermann,
Commun. Math. Phys. {\bf 97}, 211 (1985).}
\bibitem{ONA}
{R. Oehme, Prog. Theor. Phys. Suppl. {\bf 86}, 216 (1986). }
\bibitem{LJA}
{A.M. Ljapunov, `General Problems of the Stability of Motion' ,
(in Russian) (Charkov 1892); French translation in Annals
of Mathematical Studies, No.17 (Princeton University Press,
Princeton, 1947). }
\bibitem{MAL}
{I.G. Malkin, `Theory of the Stability of Motion' , (in Russian)
(Ghostekhizdat, Moscow, 1952); German translation (Oldenbourg,
M\"unchen, 1959).}
\bibitem{OSZ}
{R. Oehme, K. Sibold, and W. Zimmermann, (unpublished).}
\bibitem{BZK}
{T. Banks and A. Zaks, Nucl. Phys. {\bf B196}, 189 (1982); \\
J. Kubo,  Phys. Rev. {\bf D52}, 6475 (1995). }
\bibitem{KUS}
{S. Elizur, A. Giveon, D. Kutasov, E. Rabinovici and A. Schwimmer,
Nucl. Phys. {\bf B505}, 202 (1997); 
A. Giveon and D. Kutasov, hep-th/9802067, Rev. Mod. Phys. (to be published).}


\end{thebibliography}
\end{document}